\documentclass[twocolumn,showpacs,preprintnumbers,amsmath,amssymb,prl]{revtex4}

\usepackage{graphicx}
\usepackage{dcolumn}
\usepackage{bm}
\usepackage{epsfig}

\begin{document}

\title{The Role of Reconstructed Surfaces in the Intrinsic Dissipative Dynamics of Silicon Nanoresonators}

\author{M. Chu$^1$, R. E. Rudd$^2$, M. P. Blencowe$^1$}
\affiliation{$^{1}$Department of Physics and Astronomy, Dartmouth College,
Hanover, New Hampshire 03755, USA\\
$^2$Lawrence Livermore National Laboratory, Condensed Matter Physics Div., L-045, Livermore,  CA 94551 USA}

\begin{abstract}
Dissipation in the flexural dynamics of doubly clamped nanomechanical
bar resonators is investigated using molecular dynamics simulation.
The dependence of the quality factor ($Q$) on temperature and the size of
the resonator is calculated from direct simulation of the oscillation of a
series of Si $\langle$001$\rangle$ bars with bare $\{100\}$ dimerized
surfaces.  The bar widths range from 3.3 to 8.7~nm, all with a fixed length of 22~nm.
The fundamental mode frequencies range from 40 to 90~GHz and
$Q$ from $10^2$ near 1000~K to $10^4$ near 50~K.  
The quality factor is shown to be limited by defects in the
reconstructed surface.
\end{abstract}

\maketitle

There is currently much interest in ultrahigh frequency nanoscale mechanical
resonators for both fundamental science and engineering applications. From a
fundamental perspective, nanoelectromechanical systems (NEMS) are leading
candidates for observing macroscopic quantum
behavior~\cite{schwab05,blencowe04}, while applications include force sensing
and frequency generation~\cite{ekinci05,cleland02}. Achieving the quantum
limit requires resonators with fundamental mode frequencies $\omega$ in the
radio-to-microwave frequency range that are also well separated from other
low-lying modes on the scale of the resonance frequency width.  The frequency
width is determined by the mechanical energy damping rate $\gamma$ and hence
large quality factors  $Q=\omega / \gamma$ are required.  Force sensing and
frequency generation applications also require large quality factors.
For example, the relative uncertainty in a generated frequency is
$1/Q$~\cite{clark05}. It has also recently been demonstrated that
nanoresonators can be used as mass detectors with sensitivities approaching
the single Dalton (i.e., carbon atom mass)
level~\cite{yang06,ono03,ilic,ekinci}; when a mass
attaches to the resonator, the fundamental resonance frequency shifts and the
ability to resolve this shift depends on the quality factor.

The energy dissipation of nanomechanical resonators is described in
general by three mechanisms~\cite{roukes, foulgoc}: (1) energy losses to
the extrinsic environment through the attachment cross section of a resonator
to its support substrate (`clamping loss'), or through air friction; (2)
intrinsic  losses arising from anharmonicity (or phonon scattering),
thermoelastic damping, electron-phonon coupling and the strain-driven
motion of defects in the interior
(bulk) of a resonator; and (3) intrinsic losses arising from analogous
processes within the surface layers. For example, lattice
defects can move in response to an applied strain and then relax via
elastic radiation. The energy dissipated depends on both the microscopic
structure of the defect dynamics and on the long-wavelength acoustic
properties of the mechanical device and its surroundings. The quality factors
of the fundamental flexural and other low-lying vibrational modes have been
observed to decrease approximately linearly with decreasing resonator
size~\cite{ekinci05}. This size dependence is usually interpreted to be a
signature of surface-related dissipation mechanisms being dominant in
(sub-)micron scale resonators.

In this letter, we investigate the fundamental, intrinsic energy dissipation
for doubly clamped silicon bar nanoresonators arising from the anharmonic
interatomic potential and surface reconstruction; the resonators are in
vacuum and support atoms kept fixed so that extrinsic loss mechanisms are
suppressed.   Direct numerical simulations of the dynamics of individual atoms
are employed to examine how energy is transferred from the fundamental mode of
interest to the other modes.  We have extensively adapted the scalable,
parallel molecular dynamics (MD) code MDCASK~\cite{MDCASK}
to simulate and analyze the nanoresonator oscillation,
making use of the Stillinger-Weber (SW) potential~\cite{sw}
and the Tersoff potential~\cite{ts}
to describe the interatomic interaction.
These potentials involve nearest-neighbor two-body and
three-body interactions parameterized empirically based on
bulk elastic and thermodynamic properties of the Si diamond-cubic
crystal and, for SW, its melt point.
Both potentials are widely used and validated, achieving a reasonable
description of many properties including surface structures and
energies~\cite{balamane}.
The use of both potentials in our simulations enables their systematic
comparison in the modeling of mechanical dissipation in Si
nanoresonators, for which their relative merits are not known {\em a priori}.
Several authors~\cite{jiang04, tangney, zhao} have used MD simulations
to study energy dissipation in carbon nanotube oscillators.
Our MD study of Si resonators is conducted in a manner similar
to that used previously for the oscillations of quartz resonators~\cite{BMVK,RB}.

The simulations focus on single-crystal diamond-cubic
Si $\langle$001$\rangle$ bars with $\{$100$\}$ faces.
To accommodate the symmetric dimer $2\times1$ surface reconstruction,
square cross sections
of $m\times m$ unit cells are selected with even $m$: $m=6,8,\ldots,16$,
corresponding to widths of 3.26, 4.34, 5.43, 6.52, 7.6 and 8.69 nm. The
bar length is fixed to be 40 unit cells ($\sim 22$~nm).
Including the dimer atoms on the surface,
the number of atoms in these bars ranges from 12,360 to 84,360 atoms.
The bars are initially thermalized and mechanically equilibrated at
50~K, 100~K, 200~K, 300~K and 1000~K. The thermostat is then turned off
for the duration of the simulation: 3,000~ps (3 million time steps).
In order to clamp both ends of the bars, the positions of the atoms
initially within one unit cell at both ends of the bar are kept fixed. The
lowest flexural mode is excited by adding an initial transverse velocity in
the $x$-direction to each atom on the resonator, varying with $z$
according to the normal mode derived from linear continuum elasticity~\cite{ll}.
The initial amplitude is chosen to be $10\%$ of the bar width.

\begin{figure}[t]
\includegraphics[width=3.3in]{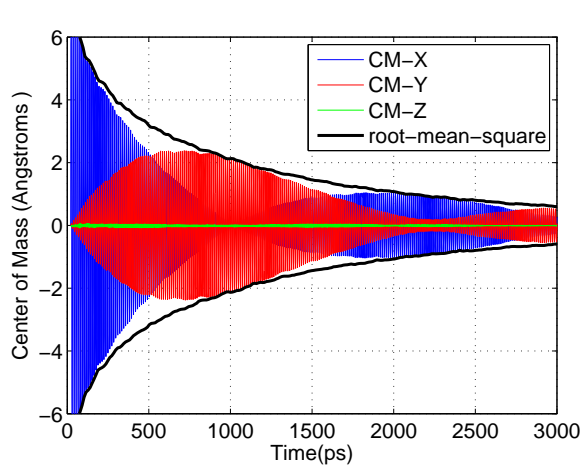}
\caption{(Color) Time evolution of the center of mass in the transverse $x$ (blue),
$y$ (red), and longitudinal $z$ (green) directions (in Angstroms) for the
8.69-nm Tersoff bar near 1000~K. The black solid curve is the root-mean-square
amplitude whose log is used to extract the damping rate.
\label{fig:beats}}
\end{figure}

\begin{figure}[t]
\includegraphics[width=3.4in]{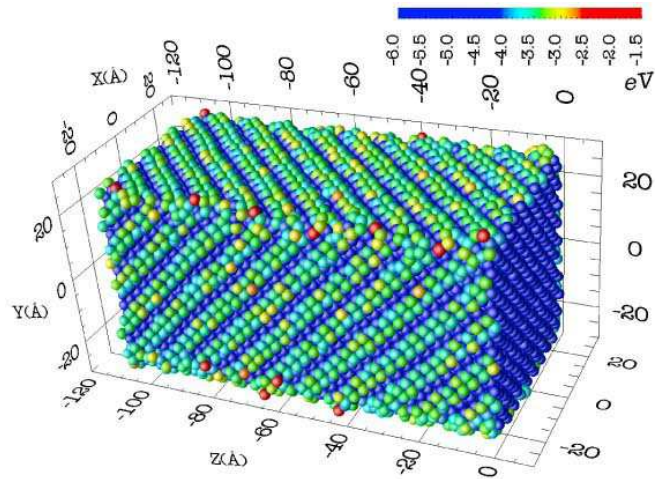}    
\caption{(Color) The left half of the 5.43-nm Tersoff bar 
near 1000~K after 3000~ps.
Atoms are colored according to their potential energy.
\label{fig:atoms}}
\end{figure}

When the bars are ``plucked" in the $x$ direction, the $x$-mode energy is
transferred partially over time into the $y$-direction mode, as evidenced by
the center of mass trajectory shown in Fig.~\ref{fig:beats}.
The energy transfer between the two transverse modes is attributed
to the fact that the silicon dimer atoms on the bar surface $\{100\}$
align themselves in the diagonal direction as shown in Fig.~\ref{fig:atoms}.
The dimerized surface of the bar breaks the cubic
symmetry of the bulk and, we infer, induces the mixing of the two transverse
modes. A tiny amount of the transverse mode energy is also transfered
into the longitudinal mode in the $z$ direction, but the transfer is very
limited due to the clamping at both ends.

\begin{figure}[t]
\includegraphics[width=3.0in]{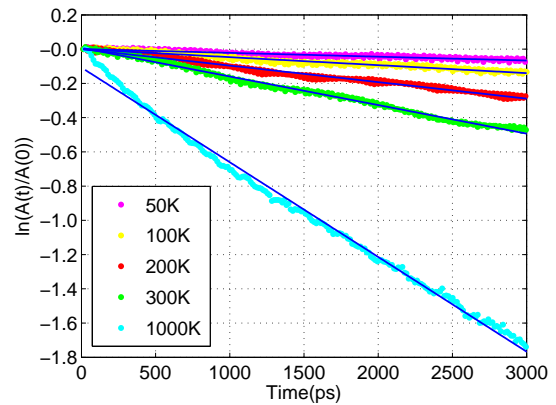}
\caption{(Color) Log of the amplitude $A(t)$ relative 
to the initial amplitude $A(0)$ versus time $t$ at various temperatures for
the 6.52-nm Tersoff bar. The dotted curves are from the
simulations and the straight solid lines in blue are the fits
whose slopes correspond to one half the energy damping rates.
\label{fig:amplitude}}
\end{figure}

The quality factor $Q$ is calculated as the ratio of the mode frequency to the
energy damping rate, $\omega/\gamma$,
where the frequencies of the dominant amplitude modes are
identified from the peaks of the Fourier transform of the center of mass
position. The frequencies for the two dominant transverse modes
are very close to each other, giving rise to
the beat pattern observed in Fig.~\ref{fig:beats}.
The damping rate is extracted as twice the slope in the time
trajectory of the log of the envelope of the amplitude which is
found to decay exponentially at a constant rate to a good approximation,
as shown in Fig.~\ref{fig:amplitude}.
Due to the mode mixing, the amplitude is taken to be the
root-mean-square of the sum of the mean-adjusted center-of-mass amplitudes in
all three directions.

\begin{figure}[t]
 \includegraphics[width=3.0in]{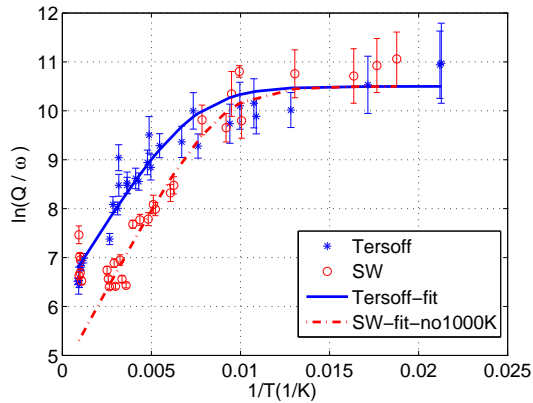}
 \caption{(Color) Temperature dependence of the damping rates, 
plotted in terms of
$\ln(Q/\omega)$ versus $1/T$ for Tersoff bars in blue and
Stillinger-Weber bars in red. $\omega$ is the mode frequency in radians/ps.
The curves are the best fits for ${\mathrm{ln}}(\gamma_1+\gamma_2)^{-1}$.
The temperature of each data point represents
an average over the whole simulation period.
The error bars are estimated from the uncertainty in the slope coefficient
for the least-squares fitting of a straight line.
\label{fig:Q}}
\end{figure}

The temperature dependence of the damping rates for the
various bars is plotted in Fig.~\ref{fig:Q}.
The fundamental mode frequencies range from 40 to 90~GHz,
increasing with bar thickness and decreasing slightly with temperature.
As the temperature increases, the quality factor decreases
from $10^4$ near 50~K down to $10^2$ near 1000~K.
The room temperature values are of the order of,
but somewhat larger than, the highest quality experimental
microresonators scaled to the simulated sizes~\cite{ekinci05}.
Quantitatively, the damping can be attributed
to an athermal process which dominates at low temperatures and
a thermal relaxation process which is manifested at room temperatures.
Denote the corresponding damping rates for the two processes to be
$\gamma_1$ and $\gamma_2$ respectively:
\begin{equation}
Q(T)=\frac{\omega(T)}{\gamma_1+\gamma_2 (T)},
\end{equation}
where $\omega$ is the mode frequency of the resonator and
the athermal process is assumed to have the constant damping rate $\gamma_1$
for both Tersoff and Stillinger-Weber bars.
$\gamma_2$ describes the thermal relaxation process
in a standard linear solid and behaves according to Arrhenius' law~\cite{defectq},
\begin{equation}
\gamma_2(T)  \approx \gamma_0 \exp (-H/k_{\mathrm B}T).
\end{equation}
where $H$ is the activation energy and $\gamma_0$ is the constant damping rate
in the high temperature limit.

The curves in Fig.~\ref{fig:Q} give the best fits of
this functional form to the two data sets.
The fitted athermal damping rate is
$\gamma_1 \approx 2.75\times 10^{-5} $ ps$^{-1}$ for both interatomic potentials.
The thermal damping rate is estimated to
have $\gamma_0 \approx 1.83\times 10^{-3}$ ps$^{-1}$
and $H\approx 50.9$~meV for the Tersoff case
while $\gamma_0 \approx 9.22\times 10^{-3}$ ps$^{-1}$ and $H\approx 57.9$~meV
for the Stillinger-Weber case.
When extracting these fitting parameters we have excluded the Stillinger-Weber
bars near 1000K where an additional relaxation process appears to take place.
We will discuss the high temperature regime in more detail below.

\begin{figure}[t]
\mbox{\begin{tabular}{l}
      \includegraphics[width=3.0in]{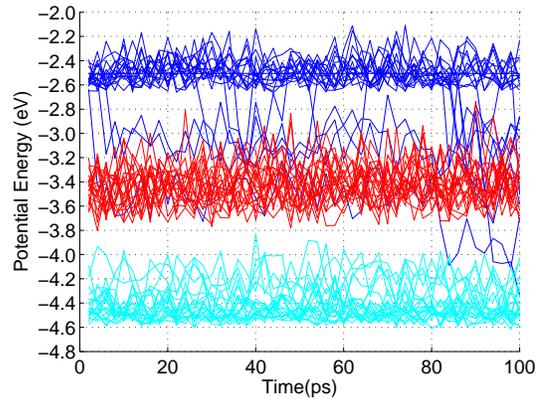}
     \end{tabular}}
\caption{(Color) Potential energy of randomly selected sample atoms in the 
5.43-nm Tersoff bar at $\sim$1000~K over the first 100 ps. The three energy 
bands, denoted as the edge (blue), the dimer (red) and the bulk (cyan), 
each comprise 20 sample atoms.}
\label{fig:PE}
\end{figure}

Over the course of the simulation various degrees of surface roughening are
observed as shown in Fig.~\ref{fig:atoms}.
In most bars the roughening on the surfaces and the edges 
is more pronounced at higher temperatures
and there is less roughening in the Tersoff bars, 
which correspondingly have less damping and
higher $Q$'s than the Stillinger-Weber bars. At higher temperatures, a greater
portion of the edge region was roughened as compared with the interior surface
and bulk regions.  This observation is consistent with theoretical and
experimental studies on edge roughening of equilibrium crystal
shapes, which can be explained by the thermal
relaxation of the steps on the surface~\cite{jeong, marchenko, shenoy}.
The activation energies extrapolated from Fig.~\ref{fig:Q} are smaller
than the activation energies for various diffusion processes
of $2\times 1$ Si dimers on the $\{100\}$ surface, ranging between
0.7~eV to 1.36~eV~\cite{borovsky}.
In fact they are comparable to step creation energies;
e.g., 10~meV/$a$ for the single-layer step, denoted as $S_A$,
or 50~meV/$a$ for the double-layer step, denoted as $D_B$, where type
$A (B)$ refers to the steps with dimer rows on the upper terrace
oriented parallel (perpendicular) to the step edge~\cite{chadi} and
$a \approx {3.85} $~\AA \hbox{\hskip 1pt} is the $1\times 1$ 
surface lattice constant.
Hence the dissipation is very likely dominated by the thermal relaxation
of the step edges and not by the diffusion of dimers.

The coloring of the atoms according to their potential energies 
in Fig.~\ref{fig:atoms} suggests that the atoms cluster into
three energy ``bands'', denoted here as the edge atoms whose energy are 
greater than -3~eV, the bulk atoms with energy less than -4~eV and
the surface dimer atoms with energy between -3~eV and -4~eV. 
Figure~\ref{fig:PE} is a sample plot of randomly selected atoms in each band. 
Most atoms remain within the same energy band throughout the simulation. 
As temperature rises, more atoms hop between bands. 
All the atoms that undergo the largest energy hopping 
are located within one unit cell or so from the geometric edges. 
The work done by the fundamental mode on the atoms during these hops
is released irreversibly and leads to dissipation.
We find that the activation energies associated with the hopping rates 
are of the same order of magnitude as those extracted from Fig.~\ref{fig:Q}.

Another observation is that the fitted damping rates in Fig.~\ref{fig:Q} 
do not seem to depend on the bar size.  This invariance
is another indication of the quality factor being limited by defects
near the geometric edge regions. When the edge defects dominate the dissipation,
different cross section but equal length bars are expected to
have similar damping rates at a given temperature.
Current experiments~\cite{ekinci05, carr, yasumura, liu, yang}, on the other hand,
involve larger silicon resonators than those in our simulations.
These resonators, usually made from a top-down approach,
are expected to have vicinal surfaces with step edges in addition to the geometrical edges.
The typical width between regularly spaced step arrays ranges
between 2-10~nm~\cite{crook, itoh}.
When the resonator dimensions are large as compared with the step spacings,
the damping rate is expected to depend on the surface area.

At room temperature the vicinal silicon surface shows a double-step reconstruction 
above a critical miscut angle of $3^o$ while at smaller miscut angles
the single-layer steps dominate the surface~\cite{alerhand, wierenga}.
For surface with a large miscut angle, a phase transition 
from $D_B$ to $S_A$ steps has been observed near 800~K~\cite{barbier}. 
Hence the smaller formation energy of $S_A$ steps may be related to
the smaller slope in Fig.~\ref{fig:Q} for the Stillinger-Weber bars near 1000K.

In conclusion, we have conducted the first systematic study of dissipation
and the resulting quality factors of nanoscale bar resonators
using molecular dynamics simulations, finding that the
quality factors of the Si nanoresonators are 
dominated by thermal relaxation of the step edges.
Furthermore, transfer of energy between modes with
polarization transverse to the $\{100\}$ surfaces is observed, due to symmetry
breaking of the bulk by the dimer surface reconstruction. The present results
further reinforce the relevance of the surface for the mechanical properties
of nanoscale resonators.

We thank Mike Geller and Makoto Itoh for useful discussions and the Research
Computing Department at Dartmouth College for their technical support. 
This work was supported in part the National Science Foundation under NIRT grant no.
CMS-0404031 and by the National Center for Supercomputing Applications (NCSA), 
utilizing Xeon Linux Cluster under Grant no. DMR050043N and 
the TERAGRID MRAC Grant no. DMR060023N.
R.E.R.'s work was performed under the auspices of the U.S.\ Dept.\ of Energy
by the Univ.\ of California, Lawrence Livermore National Laboratory, under
Contract W-7405-Eng-48.

\end{document}